\documentclass[amsthm]{elsart}

\usepackage{yjsco}
\usepackage{natbib}

\usepackage{amsmath,amstext,amssymb}
\usepackage{url}
\usepackage{hyperref}

\def\K{{\sf{K}}}
\def\R{{\sf{R}}}

\newcommand{\ZZ}{{\mathbb Z}}

\def\sO{{O {\;\!\tilde{}}}}

\newcommand{\rddots}{\mathinner{
    \mkern2mu\raise1pt\hbox{.}
    \mkern2mu\raise4pt\hbox{.}
    \mkern1mu\raise7pt\vbox{\kern7pt\hbox{.}}
    \mkern1mu}}

\begin{document}
\begin{frontmatter}

\title{Kaltofen's division-free determinant algorithm differentiated for 
matrix
 adjoint computation}

\thanks{This research was partly supported by 
the French  National Research Agency, ANR
Gecko.}

\author{Gilles Villard}
\address{CNRS, Universit\'e de Lyon, INRIA\\[0.2cm]
Laboratoire LIP, ENSL, 
46, All\'ee d'Italie, 69364 Lyon Cedex 07, France}
\ead{Gilles.Villard@ens-lyon.fr}
\ead[url]{http://perso.ens-lyon.fr/gilles.villard}


\begin{abstract}
Kaltofen has proposed a new approach in \citep{Kal92} for computing
matrix determinants
without divisions. 
The algorithm is based on a baby steps/giant steps construction of Krylov subspaces,
and computes the determinant as the constant term of a characteristic polynomial. 
For matrices over an abstract ring, by the results of \citet{BaSt82}, 
 the 
determinant algorithm, actually a straight-line program, 
leads to an algorithm with the same complexity for computing the
adjoint of a matrix.
However, the latter adjoint algorithm is obtained by the reverse mode   
of automatic differentiation, hence somehow is not ``explicit''.
We present an alternative (still closely related)  algorithm for the adjoint that 
can be implemented directly, we mean without resorting to an automatic transformation.  
The algorithm is deduced by applying program differentiation techniques ``by hand'' to Kaltofen's method, and 
is completely decribed. As subproblem, we study  the differentiation of 
programs that compute minimum polynomials of lineraly generated sequences, and we 
use a lazy polynomial evaluation mechanism  for reducing the cost 
of Strassen's avoidance of divisions in our case. 
\end{abstract}


\end{frontmatter}

%
%

\section{Introduction} \label{sec:intro}

Kaltofen has proposed in~\citep{Kal92} a new approach for computing matrix determinants.
This approach has brought breakthrough ideas for improving the complexity estimate for the problem of 
computing the determinant without divisions over an abstract ring (see \citep{Kal92,KaVi04-2}).
With these foundations, the algorithm of \citet{KaVi04-2} computes the determinant in 
$O(n^{2.7})$ additions, subtractions, and multiplications. 
The same ideas also lead to the currently 
best known bit complexity estimate of \citet{KaVi04-2} for the problem  
of computing the characteristic polynomial.

We consider the straigth-line programs of~\cite{Kal92}  
for computing the determinant over abstract fields or rings (with or without divisions).  
Using the reverse mode of automatic differentiation
(see~\cite{Lin70,Lin76}, and \citep{OWB71}), 
a straight-line program for computing the determinant 
of a matrix $A$ can be (automatically) transformed 
into a program for computing the adjoint matrix $A ^{*}$ of
$A$.  This principle, stated by~\citet[Cor.\,5]{BaSt82}, 
is also applied by~\citet[Sec.\,1.2]{Kal92} for computing $A^*$.
Since the adjoint 
program is derived by an automatic process, few is known
about the way it computes the adjoint. 
The only available information seems to be the determinant program itself, and the  
knowledge we have on the differentiation process. 
Neither the adjoint program can be described, or implemented, without 
resorting to an automatic differentiation tool. 
 
In this paper, 
by studying the differentiation of Kaltofen's determinant algorithm step by step,  
we produce an ``explicit'' adjoint algorithm. 
The determinant algorithm, that we first recall in Section~\ref{sec:detK} over 
an abstract  field~$\K$, 
uses a Krylov subspace construction, hence mainly reduces to 
vector times matrix, and matrix times matrix products. Another operation involved is
computing the minimum polynomial of a linearly generated sequence.   We apply the 
program differentiation mechanism, reviewed in Section~\ref{sec:autodiff}, 
to the different steps of the determinant program 
in Section~\ref{sec:differentiation}. This leads us to the description of a 
corresponding new adjoint program over a field, in Section~\ref{sec:adjointK}.
The algorithm we obtain somehow calls to mind the matrix factorization of~\citet[(3.4)]{Ebe97}. 
We note that our objectives are similar to Eberly's ones, whose question 
was to give an explicit inversion algorithm from the parallel determinant 
algorithm of~\citet{KaPa91}.

Our motivation for studying the differentiation and resulting adjoint algorithm,
 is the  importance of the determinant approach of~\citet{Kal92}, and~\citet{KaVi04-2},
for various complexity estimates. 
Recent advances around the determinant of polynomial or integer matrices 
(see \cite{EGV00,KaVi04-2,Sto03,Sto05}),
and matrix inversion (see~\cite{JeVi06}, and \cite{Sto08})
also justify the study of the general adjoint problem. 

For computing the determinant without divisions over a ring~$\R$, 
Kaltofen applies the avoidance of divisions 
of~\citet{Str73} to his determinant algorithm over a field.  
We apply the same strategy for the adjoint. 
From the algorithm of Section~\ref{sec:adjointK} over a field, 
we deduce an adjoint algorithm over an arbitrary ring~$\R$ in Section~\ref{sec:nodiv}. 
The avoidance of divisions involves computations with truncated power series. 
A crucial point in Kaltofen's approach is a  ``baby steps/giant steps''  
scheme for reducing the corresponding power series arithmetic cost. However,
since we use the reverse mode of differentiation, the flow of computation is modified, and 
the benefit of the baby steps/giant steps is partly lost for the adjoint. This asks 
us to introduce an early, and lazy polynomial evaluation  
strategy for not increasing the complexity estimate.

The division-free determinant algorithm of~\citet{Kal92} uses $\sO(n^{3.5})$
operations in $\R$. The adjoint algorithm we propose has essentially the same cost. 
Our study may be seen as a first step for the differentiation of the more efficient  
algorithm of~\citet{KaVi04-2}. The latter would require, in particular, to 
consider asymptotically fast matrix multiplication algorithms that are not discussed 
in what follows.    

Especially in our matrix context, 
we note that interpreting programs obtained by automatic differentiation, may have 
connections with the interpretation of programs derived using the transposition principle.
We refer for instance to the discussion of~\citet[Sec.\,6]{Kal00-2}.

\vspace*{0.6cm}

\noindent
{\bf Cost functions.}
We let ${\sf M}(n)$ be such that two univariate polynomials of degree
$n$ over
an arbitrary ring $\R$ can be multiplied using ${\sf M}(n)$ operations in~\R.
The algorithm of \citet{CaKa91} allows ${\sf M}(n)=O(n\log n
\log\log n)$. The function $O({\sf M}(n))$ also measures the cost of truncated power series 
arithmetic over $\R$.
For bounding the cost 
of polynomial gcd-type computations over a commutative field $\K$ we define the function ${\sf G}$.
Let ${\sf G}(n)$ be such that 
the extended gcd problem (see
\citep[Chap.\,11]{vzGG99}) can be solved
with ${\sf G}(n)$ operations in $\K$ 
for polynomials of degree $2n$ in $\K[x]$. 
The recursive
Knuth/Sch\"onhage half-Gcd algorithm (see \citep{Knu70,Sch71,Moe73})
allows ${\sf G}(n)=O({\sf M}(n)\log n)$.
The minimum polynomial of degree~$n$, of a linearly generated sequence given by its 
first~$2n$ terms,  can be
computed in ${\sf G}(n) +O(n)$ operations (see  \citep[Algorithm 12.9]{vzGG99}).
We will often use the notation $\sO$ that indicates missing factors
of the form $\alpha (\log n )^{\beta}$, for two 
positive real numbers $\alpha$ and $\beta$.

%
%

\section{Kaltofen's determinant algorithm over a field} \label{sec:detK}

 Kaltofen's determinant algorithm extends
the Krylov-based method of~\citet{Wie86}. 
The latter approach is successful in various situations. 
We refer especially to the algorithms of
\citet{KaPa91} and \citet{KaSa91}
around exact linear system solution that has served as basis for
subsequent works.  
We may also point out  the various questions investigated by~\citet{CEKSTV01-2}, 
and references therein.  

Let $\K$ be a commutative field. We consider $A \in \K ^{n \times n}$,
$u \in \K ^{ 1 \times n}$, and $v \in \K ^{n \times 1}$.
We introduce the Hankel matrix $H= \left(uA^{i+j-2}v\right) _{1\leq i,j \leq n} \in \K ^{n \times n}$,
and let $h_k=uA^kv$ for $0\leq k \leq 2n-1$.
We also assume that $H$ is non-singular:
\begin{equation} \label{eq:defH}
\det H = 
\det 
\left[
\begin{array}{cccc}
uv & uAv & \ldots & uA ^{n-1}v \\
uAv & uA ^{2}v & \ldots & uA ^{n}v \\
\vdots & \ddots & \vdots & \vdots\\
uA ^{n-1}v & \ldots & \ldots &  uA ^{2n-2}v
\end{array}
\right] \neq 0.
\end{equation}
In the applications, (\ref{eq:defH}) 
is ensured either by construction of $A,u$, and
$v$, as in \citep{Kal92,KaVi04-2}, 
or by randomization (see the above cited references around Wiedemann's approach, and 
\citep{Kal92,KaVi04-2}). 

One of the key ideas of \citet{Kal92} for reducing the division-free
complexity estimate for computing the determinant, is to introduce a   
``baby steps/giant steps''  
behaviour in the Krylov subspace construction. 
With baby steps/giant steps
parameters $r=\lceil 2n/s \rceil$ and $s=\lceil
\sqrt{n}\rceil$ ($rs \geq 2n$) we consider the following
algorithm.  

\begin{tabular}{rl}
~~\\[-0.16cm]
Algorithm & \!\!\!\!{\sc Det}\\
 {\em Input:} &  \!\!\!\!$A \in \K ^{n \times n}, u \in \K ^{ 1 \times n}, v \in \K ^{n \times 1}$\\
& \!\!\!\!\!\!\!\!\!
\begin{tabular}{ll} 
{\sc step i}. & \!\!\!$v_0:=v$; For $i=1,\ldots,r-1$ do $v_i := A v_{i-1}$\\[-0.16cm]
{\sc step ii}. & \!\!\!$B:=A^r$ \\[-0.16cm]
{\sc step iii}. & \!\!\!$u_0:=u$; For $j=1,\ldots,s-1$ do $u_j := u_{j-1} B$
 \\[-0.16cm]
{\sc step iv}. & \!\!\!For $i=0,1,\ldots,r-1$ do \\[-0.16cm]
&\hspace*{0.64cm}\!\!\!For $j=0,1,\ldots,s-1$ do $h_{i+jr}:=u_jv_i$  \\[-0.16cm]
{\sc step v}. & \!\!\!$f:=$ the minimum polynomial of $\{h_k\}_{0\leq k \leq 2n-1}$
\end{tabular}\\
{\em Output:} &  \!\!\!\!$\det A := (-1)^nf(0)$.  \\[-0.16cm]
~~
\end{tabular}

We ommit the proof of next theorem that establishes the correctness
and the cost of Algorithm~{\sc Det}, and refer to~\citet{Kal92}.
We may simply note that the sequence $\{h_k\}_{0\leq k \leq 2n-1}$ is linearly generated.
In addition,  if~(\ref{eq:defH})
is true, then 
the minimum polynomial $f$ of $\{h_k\}_{0\leq k \leq 2n-1}$,
the minimum polynomial of $A$, and the characteristic (monic) polynomial of
$A$ coincide. Hence $(-1)^nf(0)$ is equal to the
determinant of $A$.   
Via an algorithm that can multiply two matrices 
of $\K ^{n \times n}$ in $O(n^{\omega})$ we have:

\begin{thm} \label{theo:proofdet}
If $A\in \K ^{n \times n}$, $u\in \K ^{ 1 \times n}$, and $v\in \K
^{n \times 1}$ satisfy~(\ref{eq:defH}),
 then Algorithm~{\sc Det}
computes the determinant of $A$ in $O(n^{\omega}\log n)$ operations
in
$\K$.
\end{thm}

 For the matrix
product we may set $\omega =3$, or $\omega=2.376$
using the algorithm of~\citet{CoWi90}.
In the rest of the paper we work with a cubic matrix multiplication algorithm. 
Our study has to be generalized if fast matrix multiplication is introduced.

%
%

\section{Backward automatic differentiation} \label{sec:autodiff}

The determinant of $A\in \K^{n\times n}$ is a polynomial $\Delta$ in
$\K[a_{1,1},\ldots,a_{i,j},\ldots,a_{n,n}]$ of
the entries of $A$. 
We denote the adjoint matrix by $A ^{*}$ such that 
$A A ^{*}= A ^{*} A = (\det A) I$. 
As noticed by~\citet{BaSt82},
the entries of $A ^{*}$ satisfy
\begin{equation} \label{eq:partial}
  a_{j,i}^{*}=\frac{\partial \Delta}{\partial a_{i,j}}, 1\leq i,j
  \leq n.
\end{equation}

The reverse mode of automatic differentiation 
 allows to transform a
program which computes $\Delta$ into a program which computes all
the partial derivatives in~(\ref{eq:partial}).
Among the rich literature about the reverse mode of automatic differentiation we may refer 
to the seminal works of \citet{Lin70,Lin76} and \citet{OWB71}. For
deriving the adjoint program from the determinant program we
follow the lines of \citet{BaSt82} and  \citet{Mor85}.  

Algorithm~{\sc Det} is a straight-line program over $\K$. For a comprehensive
study of straight-line programs for instance see
\citep[Chapter 4]{BCS97}. 
We assume that the entries of $A$ are stored initially in $n^2$ 
variables $\delta_i$, $-n^2 <  i \leq 0$. 
Then we assume that the algorithm is a sequence of arithmetic
operations in $\K$, or assignments to constants of $\K$.  
Let $L$ be the number of such operations. 
We assume that the result of each instruction is stored in a new 
variable $\delta_i$, 
hence the algorithm is seen as a sequence of instructions  
\begin{equation} \label{eq:slp1}
\delta_i := \delta_j \text{~op~} \delta_k, \text{~op} \in {\{+,-,\times,\div\}}, ~-n^2 < j,k < i, 
\end{equation}
or
\begin{equation} \label{eq:slp2}
\delta_i := c, ~c \in \K,
\end{equation}
for $1\leq i \leq L$.
Note that a binary arithmetic operation~(\ref{eq:slp1}) 
where one of the operands is a constant of $\K$ can be implemented 
with the aid of~(\ref{eq:slp2}).
For any  $0\leq i \leq L$, the determinant maybe be seen as a
rational function $\Delta _i$ of $\delta_{-n^2+1}, \ldots,
\delta_{i}$, such that  
\begin{equation} \label{eq:firstinstruct}
\Delta _0  (\delta_{-n^2+1}, \ldots, \delta_{0}) = \Delta (a_{1,1},
\ldots , a_{n,n}),
\end{equation}
and such that 
the last instruction gives 
the result:
\begin{equation} \label{eq:lastinstruct}
\det A = \delta_{L} = \Delta _{L} (\delta_{-n^2+1}, \ldots, \delta_{L}).
\end{equation}

The reverse mode of automatic differentiation computes the
derivatives~(\ref{eq:partial})
in a backward recursive way, from the derivatives
of~(\ref{eq:lastinstruct}) to those of~(\ref{eq:firstinstruct}).
Using~(\ref{eq:lastinstruct}) we start the recursion with  
$$
\frac{\partial \Delta _L}{\partial \delta_L} = 1, ~ \frac{\partial \Delta
  _L}{\partial \delta_l} = 0, ~ -n^2 < l \leq L-1.
$$
Then, 
writing 
\begin{equation} \label{eq:deltaidentities}
\Delta _{i-1} (\delta_{-n^2+1}, \ldots, \delta_{i-1})= \Delta _{i} (\delta_{-n^2+1},
\ldots, \delta_{i})
= \Delta _{i}(\delta_{-n^2+1}, \ldots, g(\delta_j, \delta_k)),
\end{equation}
where $g$ is given by~(\ref{eq:slp1}) or~(\ref{eq:slp2}),
we have 
\begin{equation} \label{eq:maindiff}
\frac{\partial \Delta _{i-1}}{\partial \delta_l}
= \frac{\partial \Delta _{i}}{\partial \delta_l}
+ \frac{\partial \Delta _{i}}{\partial \delta_{i}} 
\frac{\partial g}{\partial \delta_l}, ~ -n^2 < l \leq i-1,
\end{equation}
for $1\leq i \leq L$.
Depending on $g$ several cases may be examined. For instance, for an
addition $\delta_i := g(\delta_k,\delta_j)=\delta_k + \delta_j$, (\ref{eq:maindiff}) becomes 
\begin{equation} \label{eq:diffadd}
\frac{\partial \Delta _{i-1}}{\partial \delta_k}
= \frac{\partial \Delta _{i}}{\partial \delta_k}
+ \frac{\partial \Delta _{i}}{\partial \delta_{i}}, 
~~~\frac{\partial \Delta _{i-1}}{\partial \delta_j}
= \frac{\partial \Delta _{i}}{\partial \delta_j}
+ \frac{\partial \Delta _{i}}{\partial \delta_{i}}
,
\end{equation}
with the other derivatives ($l\neq k$ or $j$) remaining unchanged. 
In  the case of a multiplication 
$\delta_i := g(\delta_k,\delta_j)=\delta_k \times \delta_j$, (\ref{eq:maindiff}) gives that the only derivatives that are
modified are  
\begin{equation} \label{eq:diffmul}
\frac{\partial \Delta _{i-1}}{\partial \delta_k}
= \frac{\partial \Delta _{i}}{\partial \delta_k}
+ \frac{\partial \Delta _{i}}{\partial \delta_{i}}\,\delta_j, 
~~~\frac{\partial \Delta _{i-1}}{\partial \delta_j}
= \frac{\partial \Delta _{i}}{\partial \delta_j}
+ \frac{\partial \Delta _{i}}{\partial \delta_{i}}\,\delta_k.
\end{equation}

We see for instance in~(\ref{eq:diffmul}), where $\delta_j$ is used for
updating the derivative with respect to $\delta_k$, that the recursion uses
intermediary results of the determinant algorithm. 
For the adjoint algorithm, we will assume that the determinant algorithm has been executed once, and 
that the $\delta_i$'s are stored in $n^2 +L$ memory
locations. 

Recursion (\ref{eq:maindiff}) gives a practical mean, and a program,
for computing the $N=n^2$ derivatives of $\Delta$ with respect to the
$a_{i,j}$'s. For any rational function $Q$ in $N$ variables
$\delta_{-N+1},\ldots , \delta_0$ 
the corresponding general statement is: 

\begin{thm} \label{theo:BaSt} [\citet{BaSt82}] Let ${\mathcal P}$ be 
a straight-line program computing $Q$ in $L$
operations in $\K$. One can derive an algorithm $\partial {\mathcal P}$
that computes $Q$ and the $N$ partial derivatives ${\partial
  Q}/{\partial \delta_l}$ in less than $5L$ operations in $\K$. 
\end{thm}

Combining Theorem~\ref{theo:BaSt} with Theorem~\ref{theo:proofdet}
gives the construction of an algorithm $\partial${\sc Det} for computing the adjoint
matrix $A ^*$ (see \cite[Corollary 5]{BaSt82}).
The algorithm can be generated automatically via an automatic
differentiation tool\footnote{We refer for instance to 
 \url{http://www.autodiff.org}}.  
However, it seems unclear how it could be programmed directly, and,
to our knowledge, it has no interpretation of its own.

%
%

\section{Differentiating the determinant  algorithm over a field} \label{sec:differentiation}

We apply the backward recursion~(\ref{eq:maindiff}) to
Algorithm~{\sc Det} 
of Section~\ref{sec:detK} for deriving the algorithm 
$\partial${\sc Det}. 
We assume that $A$ is non-singular, hence $A^*$ is non-trivial.  
By construction, 
the flow of computation for the adjoint is reversed compared to the
flow of Algorithm~{\sc Det}, therefore we start with the differentiation 
of {\sc step v}.

%

\subsection{Differentiation of the minimum polynomial constant term computation}
\label{subsec:constantterm}

At {\sc step~v}, Algorithm~{\sc Det} computes the 
minimum polynomial $f$ of the linearly generated sequence 
$\{h_k\}_{0\leq k \leq 2n-1}$.
Let $\lambda$ be the first instruction index at which all the 
$h_k$'s are known. We apply the recursion until step $\lambda$, globally, we
mean that we compute the derivatives of~$\Delta _{\lambda}$.
After the instruction $\lambda$, the determinant is viewed as
a function $\Delta
_{\text{\sc v}}$ of the
$h_k$'s only.   
Following~(\ref{eq:deltaidentities}) we have  
$$
\det (A) = \Delta _{\lambda}(\delta_{-n^2+1}, \ldots, \delta_{\lambda}) =\Delta
_{\text{\sc v}}(h_1, \ldots , h_{2n-1}). 
$$
Hence we may focus on  the derivatives $\partial \Delta
_{\text{\sc v}} / \partial h_k$, $0\leq k \leq 2n-1$, 
the remaining ones are zero. 

Using assumption~(\ref{eq:defH}) we know that 
the minimum polynomial $f$ of $\{h_k\}_{0\leq k \leq 2n-1}$ has degree
$n$, and if $f(x)=f_0 + f_1 x + \ldots + f_{n-1} x^{n-1} + x^n$,
then $f$ satisfies 
\begin{equation} \label{eq:linsysf}
H \left[
\begin{array}{c} f_0 \\ f_1 \\ \vdots \\ f_{n-1}\end{array}
\right]
= 
\left[
\begin{array}{cccc}
h_0 & h_1 & \ldots & h_{n-1} \\
h_1 & h_{2} & \ldots & h_{n} \\
\vdots & \ddots & \vdots & \vdots\\
h_{n-1} & \ldots & \ldots &  h_{2n-2}
\end{array}
\right]
\left[
\begin{array}{c} f_0 \\ f_1 \\ \vdots \\ f_{n-1}\end{array}
\right]
= - 
\left[
\begin{array}{c} h_n \\ h_{n+1} \\ \vdots \\ h_{2n-1}\end{array}
\right]
\end{equation}
see, e.g., \citep{Kal92}, or \citep[Algorithm 12.9]{vzGG99} together with
\citep{BGY80}. Applying Cramer's rule we see that 
$$
f_0 = (-1)^n 
\det \left[
\begin{array}{cccc}
h_1 & h_2 & \ldots & h_{n} \\
h_2 & h_{3} & \ldots & h_{n+1} \\
\vdots & \ddots & \vdots & \vdots\\
h_{n} & \ldots & \ldots &  h_{2n-1}
\end{array}
\right] / \det H,
$$
hence, defining $H_A=\left( uA^{i+j-1}v\right) _{1\leq i,j \leq
  n}=\left( h_{i+j-1}\right)_{1\leq i,j \leq n} \in \K ^{n\times n}$, we obtain
\begin{equation} \label{eq:quof0}
\Delta_{\text{\sc v}} = \frac{\det H_A}{\det H}.
\end{equation}
Let $\tilde{{\mathcal K}}_u$ and 
 ${\mathcal K}_v$ be the Krylov matrices 
 \begin{equation}
   \label{eq:defKu}
   \tilde{{\mathcal K}}_u = [u ^T, A ^Tu ^T, \ldots, (A ^T) ^{n-1}u ^T]^T \in \K ^{n\times n},
 \end{equation}
and 
\begin{equation}
  \label{eq:defKv}
  {\mathcal K}_v = [v, Av, \ldots, A ^{n-1}v]  \in \K ^{n\times n}.
\end{equation}
Since $H=\tilde{{\mathcal K}}_u {\mathcal K}_v$, assumption~(\ref{eq:defH}) 
implies that both  $\tilde{{\mathcal K}}_u$ and ${\mathcal K}_v$
are non-singular. Hence, using that $A$ is non-singular, we note that 
$H_A = \tilde{{\mathcal K}}_u A {\mathcal K}_v$ also is non-singular.

For differentiating~(\ref{eq:quof0}), let us first specialize~(\ref{eq:partial}) to Hankel matrices.
We denote by $(\partial \Delta /\partial a_{i,j})(H)$ the
substitution of the $a_{i,j}$'s for the entries of $H$ in 
$\partial \Delta /\partial a_{i,j}$, for 
 $1\leq i,j \leq n$.
From~(\ref{eq:partial}) we have 
$$
h ^* _{j,i} = \frac{\partial \Delta}{\partial a_{i,j}}(H), 1\leq i,j \leq n.
$$
Since the entries of $H$ are constant along the anti-diagonals, 
we deduce that 
$$
\frac{\partial \det H}{\partial h_k} = \sum _{i+j-2=k}
\frac{\partial \Delta}{\partial a_{i,j}}(H)
= \sum _{i+j-2=k} h ^* _{j,i}= \sum _{i+j-2=k} h ^* _{i,j}, ~0 \leq
k \leq 2n-2. 
$$ 
In other words, we may write
\begin{equation}\label{eq:partialH}
\frac{\partial \det H}{\partial h_k} =  \sigma_k (H ^*),  ~0 \leq
k \leq 2n-1,
\end{equation}
where, for a matrix $M=(m_{ij})$, we define 
$$\sigma_k(M)=
0+\sum_{i+j-2=k} m_{ij}, ~1 \leq i,j \leq n.$$ 
The function 
$\sigma_k(M)$ is the sum of the entries in the anti-diagonal 
of $M$ starting with $m_{1,k+1}$
if $0 \leq k \leq n-1$, and $m_{k-n+2,n}$
if $n \leq k \leq 2n-2$.
Shifting the entries of $H$ for obtaining $H_A$ we also have 
\begin{equation}\label{eq:partialHA}
\frac{\partial \det H_A}{\partial h_k} =  \sigma_{k-1} (H_A ^*), ~0 \leq
k \leq 2n-1. 
\end{equation}
Now, differentiating~(\ref{eq:quof0}), together
with~(\ref{eq:partialH}) 
and~(\ref{eq:partialHA}), leads to 
$$
\frac{\partial \Delta_{\text{\sc v}}}{\partial h_k}
= \frac{(\partial \det H_A / \partial h_k)}{\det H} - 
\frac{(\partial \det H / \partial h_k)}{\det H} \frac{\det H_A}{\det
  H}
= \frac{(\partial \det H_A / \partial h_k)}{\det H_A} \frac{\det H_A}{\det H}
- \sigma _k (H ^{-1})  \Delta_{\text{\sc v}}
$$
and,  consequently, to 
\begin{equation} \label{eq:diff5}
\frac{\partial \Delta_{\text{\sc v}}}{\partial h_k}
= \left( \sigma _{k-1} (H_A ^{-1}) - \sigma _k (H ^{-1})\right)
\Delta_{\text{\sc v}}, ~0 \leq
k \leq 2n-1.
\end{equation}

With~(\ref{eq:diff5}) we identify the problem solved by the first step of the $\partial${\sc
  Det}
algorithm, and provide first informations for interpreting or
implementing the adjoint program.   
Various algorithms may be used for computing the minimum polynomial
(for instance see \citep[Algorithm 12.9]{vzGG99}), that will lead to
corresponding algorithms for computing the  
left sides in~(\ref{eq:diff5}).
However, we will not discuss these aspects, 
since the associated costs are not dominant in the overall
complexity.  

We have recalled, in the introduction, that the minimum polynomial 
$f$ (its constant term $f(0)$) can be computed from the $h_k$'s in ${\sf G}(n) +O(n)$
operations in $\K$. Hence Theorem~\ref{theo:BaSt}
gives an algorithm for computing the derivatives
using $5{\sf G}(n) +O(n)$ operations. Alternatively, 
in the Appendix we propose a direct approach that takes advantage of~(\ref{eq:diff5}).  
Proposition~\ref{prop:computsigma}
shows that if $f$, $H$, and $H_A$ are given, then the
${\partial \Delta_{\text{\sc v}}}/{\partial h_k}$'s
can be computed in ${\sf G}(n) + O({\sf M}(n))$ operations in~$\K$.

%

\subsection{Differentiation of the dot products}

For differentiating {\sc step~iv}, $\Delta$ is seen as a function
$\Delta _{\text{\sc iv}}$ of
the $u_j$'s and $v_i$'s. The entries of $u_j$ are used for computing 
the $r$ scalars $h_{jr},h_{1+jr}, \ldots, h_{(r-1)+jr}$ for $0\leq j
\leq s-1$.
 The entries of $v_i$ are involved in the computation of 
the $s$ scalars $h_i, h_{i+r}, \ldots, h_{i+(s-1)r}$ for $0\leq i \leq r-1$.

In~(\ref{eq:maindiff}), the new derivative ${\partial \Delta
_{i-1}}/{\partial \delta_l}$ is obtained
by adding the current instruction contribution to the 
previously computed derivative  ${\partial \Delta
_{i}}/{\partial \delta_l}$. Since 
all the $h_{i+jr}$'s are computed independently according to  
$$
h_{i+jr} = \sum _{l=0}^n (u_j)_l (v_i)_l,
$$
it follows that the derivative of $\Delta _{\text{\sc iv}}$ with respect to an
entry
$(u_j)_l$ or $(v_i)_l$
is obtained by summing up the contributions of the multiplications
$(u_j)_l (v_i)_l$. 
We obtain 
\begin{equation} \label{eq:tmpDu}
\frac{\partial \Delta _{\text{\sc iv}}}{\partial (u_j)_l} =
\sum _{i=0}^{r-1} \frac{\partial \Delta _{\text{\sc v}}}{\partial
  h_{i+jr}} (v_i)_l, ~0\leq j
\leq s-1,~1\leq l \leq n,
\end{equation}
and 
\begin{equation} \label{eq:tmpDv}
\frac{\partial \Delta _{\text{\sc iv}}}{\partial (v_i)_l} =
\sum _{i=0}^{s-1} \frac{\partial \Delta _{\text{\sc v}}}{\partial
  h_{i+jr}} (u_j)_l, ~0\leq i
\leq r-1, ~1\leq l \leq n.
\end{equation}

By abuse of notations (of the sign $\partial$), we let $\partial u_j$ be the $n\times 1$
vector, respectively $\partial v_i$ be the $1\times n$
vector, whose entries are the
derivatives of $\Delta _{\text{\sc iv}}$ with respect to the entries of $u_j$,
respectively $v_i$. Note that because of the index transposition  
in~(\ref{eq:partial}), it is convenient, here and in the following, to take 
the transpose form (column versus row) for the derivative vectors.
Defining also 
$$
\partial H = \left( \frac{\partial \Delta _{\text{\sc v}}}{\partial
  h_{i+jr}}
\right)_{
0\leq i \leq r-1, ~0\leq j \leq s-1} \in \K ^{r \times s},
$$
we deduce, from~(\ref{eq:tmpDu}) and~(\ref{eq:tmpDv}),
 that 
\begin{equation} \label{eq:diff4u}
\left[
\partial u_0, \partial u_1, \ldots, \partial u_{s-1}
\right]
=  
\left[
v_0, v_1, \ldots, v_{r-1}
\right]
 \partial H
\in \K ^{n \times s}.
\end{equation}
and 
\begin{equation} \label{eq:diff4v}
\left[\begin{array}{c}
~~~~\partial v_0~~~~\\
\partial v_1\\
\vdots\\
\partial v_{r-1}
\end{array}
\right]
= \partial H
\left[\begin{array}{c}
~~~~u_0~~~~\\
u_1\\
\vdots\\
u_{s-1}
\end{array}
\right] \in \K^{r\times n}.
\end{equation}

Identities~(\ref{eq:diff4u}) and~(\ref{eq:diff4v}) give the second step of the adjoint
algorithm. 
In Algorithm~{\sc Det}, {\sc step~iv} costs essentially $2rsn$
additions and multiplications in $\K$.
Here we have essentially $4rsn$ additions and multiplications using 
basic loops (as in  {\sc step~iv}) for calculating the matrix
products, we mean without an asymptotically fast matrix multiplication
algorithm.

%

\subsection{Differentiation of the matrix times vector and matrix products}

The recursive process for differentiating {\sc step iii} to {\sc
  step i} may be written in terms of the differentiation of the basic operation 
(or its transposed operation) 
\begin{equation} \label{eq:pq}
q := p \cdot M  \in \K^{1\times n},
\end{equation}
where $p$ and $q$ are row vectors of dimension $n$, and $M$ is an
$n\times n$ matrix. 
We assume at this point (by construction of the recursion) that column vectors 
$\partial p$ and $\partial q$ of derivatives of the determinant with respect to the entries
of $p$ and $q$, are available. 
For instance, for differentiating {\sc step~iii}, we will consider the 
$\partial u_j$'s.
We also assume that an $n \times n$ 
matrix $\partial M$, whose transpose gives the derivatives with respect to the
$m_{ij}$'s, has been computed. Initially, for  {\sc step~iii}, we will take 
$\partial B=0$.

Following the lines of previous section for obtaining~(\ref{eq:diff4u})
and~(\ref{eq:diff4v}),
we see that differentiating~(\ref{eq:pq}) amounts to updating 
$\partial p$ and $\partial M$ according to 
\begin{equation} \label{eq:diffpq}
\left\{\begin{array}{l}
\partial p := \partial p + M \cdot \partial q \in \K ^n,\\
\partial M := \partial M + \partial q \cdot p \in \K^{n\times n}.
\end{array} \right.
\end{equation}
Starting from the values of the $\partial u_j$'s computed
with~(\ref{eq:diff4u}), and from $\partial B=0$, 
 for the differentiation of {\sc step
  iii}, (\ref{eq:diffpq}) gives
\begin{equation}\label{eq:diff3}
\left\{
\begin{array}{l}
\partial u_{j-1} := \partial u_{j-1} + B \cdot \partial u_j,\\
\partial B := \partial B + \partial u_j \cdot u_{j-1}, ~j=s-1, \ldots,
1.
\end{array}
\right.
\end{equation} 

For {\sc step ii}, we mean $B:=A^r$, we show that the backward
recursion leads to 
\begin{equation}\label{eq:diffpow}
\partial A :=  \sum _{k=1}^r A ^{r-k} \cdot \partial B \cdot A ^{k-1}.
\end{equation}
Here, the notation $\partial A$ stands for the $n\times n$
matrix whose transpose gives the derivatives 
$\partial  \Delta _{\text{\sc ii}}/{\partial a_{i,j}}$.
We may show~(\ref{eq:diffpow}) 
by induction on $r$. For $r=1$, $\partial A = \partial B$ is true. 
If~(\ref{eq:diffpow}) is true for $r-1$, then let 
$C=A ^{r-1}$ and $B=CA$. Using~(\ref{eq:diffpq}), 
and overloading  the notation $\partial A$, 
we have 
$$
\left\{\begin{array}{l}
\partial C =  A \cdot \partial B \in \K ^{n \times n},\\
\partial A = \partial B \cdot C \in \K^{n\times n}.
\end{array} \right.
$$
Hence, using~(\ref{eq:diffpow}) for $r-1$,
we establish that 
$$
\begin{array}{ll}
\partial A & = \partial A + \sum _{k=1}^{r-1} A ^{r-k-1} \cdot
\partial C \cdot A ^{k-1},\\
& = \partial B \cdot C + \sum _{k=1}^{r-1} A ^{r-k-1} \cdot
( A \cdot \partial B) \cdot A ^{k-1}\\
& = \partial B \cdot A ^{r-1} + \sum _{k=1}^{r-1} A ^{r-k} \cdot
\partial B \cdot A ^{k-1} = \sum _{k=1}^r A ^{r-k} \cdot \partial B \cdot A ^{k-1}.
\end{array}
$$

Any specific 
approach for computing $A ^r$ will lead to an associated program for
computing $\partial A$. 
Let us look, in particular, at the case where {\sc step ii} of Algorithm {\sc Det} is implemented by
repeated squaring, in essentially $\log_2 r$
matrix products. Consider the recursion 
$$
\begin{array}{l}
A_0:=A\\
  \text{For~} k=1,\ldots,\log_2 r \text{~do~}  A_{2^k}:= A_{2^{k-1}} \cdot A_{2^{k-1}}\\
B:=A_r
\end{array}
$$
that computes $B:= A ^r$. The associated program for computing the
derivatives is 

\begin{equation}\label{eq:diffpowlog}
  \begin{array}{l}
    \partial A_r:= \partial B \\
 \text{For~} k=\log_2 r,\ldots, 1 \text{~do~}  \partial
A _{2^{k-1}}:= 
A _{2^{k-1}}  \cdot \partial A_{2^{k}} + \partial A_{2^{k}} \cdot A _{2^{k-1}} \\
\partial A:= \partial A_0,
  \end{array}
\end{equation}
and costs essentially $2 \log_2 r$
matrix products.

From the values of the $\partial v_i$'s
computed with~(\ref{eq:diff4v}),
 we finally 
differentiate {\sc step~i}, and update 
$\partial A$ according to 
\begin{equation}\label{eq:diff1}
\left\{
\begin{array}{l}
\partial v_{i-1} := \partial v_{i-1} + \partial v_i \cdot A,\\
\partial A := \partial A + v_{i-1}\cdot \partial v_i, ~i=r-1, \ldots,
1.
\end{array}
\right.
\end{equation}

Now, $\partial A$ is the $n\times n$
matrix whose transpose gives the derivatives 
$\partial  \Delta _{\text{\sc i}}/{\partial a_{i,j}}
= \partial  \Delta /{\partial a_{i,j}}$,
hence from~(\ref{eq:partial})  we know that 
$A ^{*} = \partial A$. 

{\sc step iii} and {\sc step i}
both cost essentially $r$ ($\approx s$) matrix times vector products. 
From  (\ref{eq:diff3}) and (\ref{eq:diff1})
the differentiated steps both require 
$r$ matrix times vector products, and $2rn^2 +O(rn)$ 
additional operations in~$\K$.

%
%

\section{The adjoint algorithm over a field}  \label{sec:adjointK}

We call {\sc Adjoint} the algorithm obtained from the
successive
differentiations of
Section~\ref{sec:differentiation}. Algorithm~{\sc Adjoint}
is detailed below. We keep the notations of previous sections.
We use in addition $U \in \K^{s\times n}$ and $V \in \K^{n\times
  r}$
(resp.  $\partial U \in \K^{n\times s}$ and $\partial V \in \K^{r\times
  n}$)
for the right sides (resp. the left sides) of (\ref{eq:diff4u}) and  (\ref{eq:diff4v}).

The cost of {\sc Adjoint} is dominated by {\sc step iv}$^*$,
which is the differentiation of the matrix power computation. 
As we have seen with~(\ref{eq:diffpowlog}), the number of operation is essentially twice as much 
as for Algorithm {\sc Det}. The code we give allows an easy implementation.

We note that if the product by $\det A$ is avoided in {\sc step i}$^*$, then the algorithm computes 
the matrix inverse $A^{-1}$. 
We may  put this into perspective with the algorithm 
given by~\citet{Ebe97}. With $\tilde{{\mathcal K}}_u$ and 
 ${\mathcal K}_v$ the Krylov matrices of~(\ref{eq:defKu}) and~(\ref{eq:defKv}), 
Eberly has proposed a processor-efficient inversion algorithm based 
on 
\begin{equation}
  \label{eq:inverseeb}
  A ^ {-1}= {\mathcal K}_v H_A^{-1} \tilde{{\mathcal K}}_u.
\end{equation}
To see whether a baby steps/giant steps version of~(\ref{eq:inverseeb}) would 
lead to an algorithm similar to {\sc Adjoint} deserves 
further investigations.

\begin{tabular}{rl}
~~\\[-0.16cm]
Algorithm & \!\!\!\!{\sc Adjoint} ($\partial${\sc Det})\\
 {\em Input:} &  \!\!\!\!$A \in \K ^{n \times n}$ non-singular, and the
 intermediary data of Algorithm {\sc Det}\\
&  \!\!\!\!All the derivatives are initialized to zero\\
& \!\!\!\!\!\!\!\!\!
\begin{tabular}{ll} 
{\sc step i}$^*$. & {\em \!\!\!/* Requires the Hankel matrices $H$ and $H_A$, see~(\ref{eq:diff5}) */} \\[-0.16cm]
 & \!\!\!${\partial \Delta_{\text{\sc v}}}/{\partial h_k}
:= \left( \sigma _{k-1} (H_A ^{-1}) - \sigma _k (H ^{-1})\right)
\det A, ~0 \leq
k \leq 2n-1$\\
{\sc step ii}$^*$. & {\em \!\!\!/* Requires the $u_j$'s and $v_i$'s, see~(\ref{eq:diff4u})
  and~(\ref{eq:diff4v}) */} \\[-0.16cm]
& \!\!\!$\partial U := V \cdot \partial
H$\\[-0.16cm]
 & \!\!\!$\partial V := \partial
H \cdot U$\\
{\sc step iii}$^*$. & {\em \!\!\!/* Requires $B=A ^r$, see~(\ref{eq:diff3}) */} \\[-0.16cm]
& \!\!\!For $j=s-1,  \ldots, 1$ do \\[-0.16cm]
&\hspace*{0.64cm}\!\!\!$\partial u_{j-1} := \partial u_{j-1} + B \cdot \partial u_j$\\[-0.16cm]
&\hspace*{0.64cm}\!\!\!$\partial B := \partial B + \partial u_j
\cdot u_{j-1}$\\
{\sc step iv}$^*$. & {\em \!\!\!/* Requires the powers of $A$, see~(\ref{eq:diffpow})
  or~(\ref{eq:diffpowlog}) */} \\[-0.16cm]
& \!\!\!$A ^* :=  \sum _{k=1}^r A ^{r-k} \cdot \partial B
\cdot A ^{k-1}$ \\
{\sc step v}$^*$. & {\em \!\!\!/* See~(\ref{eq:diff1}) */} \\[-0.16cm]
& \!\!\!For $i=r-1,  \ldots, 1$ do \\[-0.16cm]
&\hspace*{0.64cm}\!\!\!$\partial v_{i-1} := \partial v_{i-1} + \partial v_i \cdot A$\\[-0.16cm]
&\hspace*{0.64cm}\!\!\!$ A ^* :=  A ^* + v_{i-1}\cdot \partial v_i$
\end{tabular}\\
{\em Output:} &  \!\!\!\!The adjoint matrix $A ^* \in \K^{n\times n}$.  \\[-0.16cm]
~~
\end{tabular}

%
%

\section{Application to computing the adjoint without divisions} \label{sec:nodiv}

Now let $A$ be an $n\times n$ matrix over an abstract ring $\R$.   
Kaltofen's algorithm for computing the determinant of $A$ without divisions
applies Algorithm {\sc Det} on a well chosen univariate polynomial matrix $Z(z) =
C + z (A-C)$ where $C \in \ZZ ^{n \times n}$, with a dedicated 
choice of projections $u=\varphi \in \ZZ ^{1\times n}$ and $v=\psi \in \ZZ ^{n \times 1}$. 
The algorithm uses 
  Strassen's 
avoidance of divisions (see \citep{Str73,Kal92}). Since 
the determinant of $Z$ is a polynomial of degree $n$ in $z$, 
the arithmetic operations over $\K$ in {\sc Det} may be replaced by operations on
power series in $\R [[z]]$ modulo $z^{n+1}$. 
Once the determinant of $Z(z)$ is computed, the evaluation $(\det Z)(1) = \det (C +
1 \times (A-C))$ gives the determinant of $A$. 
The choice of $C, \varphi$ and $\psi$ is such that, 
whenever a division by a truncated power series is performed 
the constant coefficients are  
$\pm 1$. Therefore the algorithm necessitates no divisions. 
Note that, by construction of $Z(z)$, the constant terms of the power series 
involved when {\sc Det} is called with inputs $Z(z), \varphi$ and $\psi$,  
are the intermediary values computed by {\sc Det}
with inputs $C, \varphi$ and $\psi$.

The cost for computing the determinant of $A$ without divisions
is then deduced as follows. 
In {\sc step i} and {\sc step ii} of Algorithm {\sc Det} applied to $Z(z)$, the 
vector and matrix entries are polynomials of degree $O(\sqrt{n})$. 
The cost of {\sc step ii} dominates, and is $O(n^3 {\sf M}(\sqrt{n}) \log n)=
\sO(n ^3 \sqrt{n})$ operations in~$\R$.
{\sc step iii}, {\sc iv}, and {\sc v} cost $O(n^2 \sqrt{n})$ operations on power series 
modulo $z^{n+1}$, that is $O(n^2{\sf M}(n)\sqrt{n})$ operations in~$\R$.
Hence $\det Z(z)$ is computed in $\sO(n^3\sqrt{n})$ operations in~$\R$, 
and $\det A$ is obtained with the same cost bound. 

An main property of Kaltofen's approach (which also 
holds for the improved blocked version of~\citet{KaVi04-2}), 
is that the scalar value $\det A$ 
is obtained via the computation 
of the polynomial value $\det Z (z)$.
This property seems to be lost with the adjoint computation. 
We are going to see how Algorithm 
{\sc Adjoint} applied to
$Z(z)$ allows to compute $A ^{*} \in \R ^{n\times n}$ in time $\sO(n^3\sqrt{n})$ operations in~$\R$,
but does not seem to allow the computation of $Z ^{*}(z) \in \R[z] ^{n\times n}$ 
with the same complexity estimate.
Indeed, a key point in Kaltofen's approach 
for reducing the overall
complexity estimate, is to compute with small degree polynomials (degree $O(\sqrt{n})$) in  
{\sc step i} and {\sc step ii}. However, 
since the adjoint algorithm has
a reversed flow, this point does not seem to be relevant 
for {\sc Adjoint}, where polynomials of degree $n$ are involved from the beginning. 

Our approach  for computing $A^*$ over~$\R$ keeps the idea of running 
Algorithm {\sc Adjoint} with input $Z(z)=C+z(A-C)$, such that 
$Z^*(z)$ has degree less than $n$, and gives $A^*=Z^*(1)$. In Section~\ref{subsec:divifree}, we verify 
that the implementation 
using Proposition~\ref{prop:computsigma}, needs no divisions.
We then show in  Section~\ref{subsec:lazy} how to establish the cost estimate $\sO(n^3\sqrt{n})$.
The principle we follow is to start  
evaluating polynomials at $z=1$ as soon as computing with the entire polynomials is 
prohibitive.  

%

\subsection{Division-free Hankel matrix inversion and anti-diagonal sums}
\label{subsec:divifree}

In Algorithm {\sc Adjoint}, divisions may only occur during the 
anti-diagonal sums computation. We verify here that with the matrix $Z(z)$,
and the special projections $\varphi \in \ZZ ^{1 \times n},\psi \in \ZZ^{n \times 1}$, the approach 
described in the Appendix   
for computing the anti-diagonal sums requires no divisions. 
Equivalently, since we use Strassen's avoidance of divisions, 
we verify that with the matrix $C$ and the projections 
$\varphi,\psi$, the approach necessitates no divisions. 
As we are going to see, this a direct consequence of the construction of~\citet{Kal92}.

Here we let $h_k = \varphi C^k \psi$ for $0 \leq k \leq 2n-1$,   
$a(x)=x^{2n}$, and $b(x)=h_0 x^{2n-1}+h_1 x^{2n-2}+ \ldots + h_{2n-1}$.
The extended Euclidean scheme with inputs $a$ and $b$ 
leads to a normal sequence, and after $n-1$ and $n$ steps of the scheme,  we get
(see~~\citep[Sec.\,2]{Kal92}): 
\begin{equation}
  \label{eq:euclidg}
s(x)a(x) +  t(x) b(x) = c(x), \text{with}\, \deg  s= n-2, \deg t=n-1, \deg c =n, 
\end{equation}
and
\begin{equation}
  \label{eq:euclidf}
\bar{s}(x)a(x) +  \bar{t}(x) b(x) = \bar{c}(x), \text{with}\, \deg  \bar{s}= n-1, \deg \bar{t}=n, 
\deg \bar{c}=n-1. 
\end{equation}
The polynomial $\bar{t}$ is such that 
\begin{equation}
  \label{eq:bart}
\bar{t}=\pm x^n + \text{intermediate~monomials~} + 1 = \pm f,
\end{equation}
with $f$ the minimum polynomial of $\{h_k\}_{0\leq k \leq 2n-1}$.
One may check, in particular, that the $n$ equations obtained by identifying the coefficients of 
degree $2n-1 \geq k \geq n$ in~(\ref{eq:euclidf})
give the linear system~(\ref{eq:linsysf}), that defines~$f$. 
The polynomial $c$ also has leading coefficient $\pm 1$. By identifying 
the coefficients of 
degree $2n-1 \geq k \geq n$ in~(\ref{eq:euclidg}), 
we obtain: 
\begin{equation} \label{eq:linsysg}
H \left[
\begin{array}{c} t_0 \\ t_1 \\ \vdots \\ t_{n-1}\end{array}
\right]
= 
\left[
\begin{array}{cccc}
h_0 & h_1 & \ldots & h_{n-1} \\
h_1 & h_{2} & \ldots & h_{n} \\
\vdots & \ddots & \vdots & \vdots\\
h_{n-1} & \ldots & \ldots &  h_{2n-2}
\end{array}
\right]
\left[
\begin{array}{c} t_0 \\ t_1 \\ \vdots \\ t_{n-1}\end{array}
\right]
= \pm
\left[
\begin{array}{c} 0 \\ 0 \\ \vdots \\ 1\end{array}
\right].
\end{equation}
Therefore $t=\pm g$ with $g$ the polynomial needed for computing~(\ref{eq:sigmaklowH})-(\ref{eq:sigmakhighHA}), 
in addition to $f$. Since $C, \varphi$, and $\psi$ are such that the extended Euclidean scheme 
necessitates no divisions (see~~\citep[Sec.\,2]{Kal92}), we see that both $f$ and $g$ may be computed 
with no divisions. The only remaining division in the algorithm for 
Proposition~\ref{prop:computsigma} is at~(\ref{eq:lastcolHA}). 
From~(\ref{eq:bart}), this division is by $f_0=1$.

%

\subsection{Lazy polynomial evaluation and division-free adjoint computation}
\label{subsec:lazy}

We run Algorithm {\sc Adjoint} with input $Z(z) \in \R[z]^{n\times n}$,
and start with operations on truncated power series modulo $z^{n+1}$.
We assume that Algorithm {\sc Det} has been executed, and that its 
intermediary results have been stored. 

Using Proposition~\ref{prop:computsigma} and previous section, 
{\sc step i}$^*$ requires $O({\sf G}(n){\sf M}(n))= \sO(n^2)$ operations 
in~$\R$ for computing $\partial H(z)$ of degree $n$ in $\R [z]^{r\times s}$.
{\sc step ii}$^*$, {\sc step iii}$^*$, and {\sc v}$^*$ cost $O(n^2 \sqrt{n})$
operations in~$\K$, hence, taking into account the power series operations,
 this gives $O(n^2 {\sf M}(n) \sqrt{n}) = \sO(n^3 \sqrt{n})$
operations in~$\R$ for the division-free version. 
The cost analysis of {\sc step iv}$^*$, using~(\ref{eq:diffpowlog}) over power series modulo $z^n$, 
leads to $\log_2 r$ matrix products, hence to the time bound $\sO(n^4)$, greater than the target estimate $\sO(n^3\sqrt{n})$. As noticed previously, {\sc step iii}
of Algorithm {\sc Det} only involves polynomials of degree $O(\sqrt{n})$, 
while the reversed program for {\sc step iv}$^*$ of Algorithm {\sc Adjoint},
relies on $\partial B(z)$ whose degree is~$n$.

Since only $Z^*(1)=A^*$ is needed, our solution, for restricting the cost to $\sO(n^3\sqrt{n})$,
 is to start evaluating at 
$z=1$ during {\sc step iv}$^*$. However, since power series multiplications are done 
modulo $z^n$, this evaluation must be lazy. The fact that 
matrices $Z^k(z)$, $1\leq k \leq r-1$, of degree at most $r-1$ are involved, enables the following.  
Let $a$ and $c$ be two polynomials such that $\deg a + \deg c = r-1$ in $\R[z]$,  
and let $b$ be of degree $n \geq r-2$ in $\R[z]$.
Considering the highest degree part of $b$, and evaluating the lowest degree part at $z=1$, 
we define  
$b_H(z) = b_nz^{r-2} + \ldots + b_{n-r+2} \in \R[z]$ and 
$b_L = b_{n-r+1} + \ldots  +b_0 \in \R$. 
We then remark that 
\begin{equation}
  \label{eq:lazyeval}
\begin{array}{ll}
\left( a(z)b(z)c(z) \bmod z^{n+1}\right)(1) &= \left( a(z)(b_H(z)z^{n-r+2} +b_L)c(z) \bmod z^{n+1}\right)(1), \\
 &= \left( a(z)b_H(z)c(z) \bmod z^{r-1}\right)(1) +  \left( a(z)b_Lc(z)\right)(1).
\end{array}
\end{equation}
For modifying
 {\sc step iv}$^*$, we follow the definition of $b_H$ and $b_L$, 
and first compute $\partial B_H(z) \in \R[z]^{n \times n}$ of degree $r-2$, 
and $\partial B_L \in \R^{n\times n}$. Applying~(\ref{eq:lazyeval}), the sum 
$\sum _{k=1}^r Z ^{r-k}(z) \cdot \partial B (z) \cdot Z^{k-1}(z)$ may then be evaluated at 
$z=1$ by the program
\begin{equation}
  \label{eq:modifstep4}
  \begin{array}{ll}
  \text{Modified {\sc step iv}}^*.~~ &   Z^* := \left( \sum _{k=1}^r Z ^{r-k}(z) \cdot \partial B_H(z)
\cdot Z ^{k-1}(z) \bmod z^{r-1} \right) (1)\\
&   Z^* := Z^* + \left( \sum _{k=1}^r Z ^{r-k}(z) \cdot \partial B_L
\cdot Z ^{k-1}(z) \right) (1), 
  \end{array}
\end{equation}
in $\sO(n^3 {\sf M}(r)) = \sO(n^3\sqrt{n})$
operations in~$\R$.
This leads to an intermediary value $Z^* \in \R ^{n\times n}$ before {\sc step v}$^*$.  
The value is updated at {\sc step v}$^*$ with power series operations, 
and a final evaluation at $z=1$ in time $\sO(n^2r {\sf M}(n))=\sO(n^3\sqrt{n})$. 
Since only {\sc step iv}$^*$
has been modified, we obtain the following result. 
\begin{thm}
Let $A \in \R ^{n \times n}$. If  
Algorithm {\sc Adjoint}, modified according to~(\ref{eq:modifstep4}), 
is executed with input $Z(z)=C+z(A-C)$, power series operations 
modulo $z^{n+1}$, and a final evaluation at $z=1$, then 
the matrix adjoint $A^*$ is computed in $\sO(n^3\sqrt{n})$
operations in~$\R$.
\end{thm}

%
%

\section{Concluding remarks}

We have developed an explicit algorithm for computing the matrix 
adjoint using only ring arithmetic operations. The algorithm has 
complexity estimate $\sO(n^{3.5})$. It represents a practical alternative 
to previously existing solutions for the problem, that rely  
on automatic differentiation of a determinant algorithm. Our description of the 
algorithm allows direct implementations. It should help understanding how the    
adjoint is computed using Kaltofen's  baby steps/giant steps construction.
 Still, a full mathematical explanation 
deserves to be investigated.  
Our work  has to be generalized 
 to the 
 block algorithm of~\citet{KaVi04-2} (with the use of fast matrix multiplication algorithms)
 whose complexity estimate 
is currently the best known for computing the determinant, and 
the adjoint without divisions. \\

\noindent
{\bf Acknowledgements.} We thank Erich Kaltofen who has brought reference~\cite{OWB71} to our attention.

%
%

\section*{Appendix: Hankel matrix inversion and anti-diagonal sums}

For implementing~(\ref{eq:diff5}), 
we study the computation 
of the anti-diagonal sums $\sigma _k$
of $H^{-1}$ and $H_A^{-1}$.

We first use the formula of~\citet{LaChCa90} for Hankel 
matrices inversion.
The minimum polynomial $f$ of $\{h_k\}_{0\leq k \leq 2n-1}$ is 
$f(x)=f_0 + f_1 x + \ldots + f_{n-1} x^{n-1} + x^n$, 
and satisfies~(\ref{eq:linsysf}).
Let the last column of $H^{-1}$ be given 
by 
\begin{equation} \label{eq:linsyslastcol}
H \,[g_0, g_1, \ldots, g_{n-1}]^T = [0, \ldots, 0, 1]^T
\in \K ^n. 
\end{equation}
Applying~\citep[Theorem\,3.1]{LaChCa90}  
with~(\ref{eq:linsysf}) and~(\ref{eq:linsyslastcol}),
we know that 
\begin{equation} \label{eq:invH}
 H^{-1}= \left[
      \begin{array}{cccc}
        f_{1} & \kern-3pt\ldots & f_{n-1} & \kern2pt 1 \\
        \vdots &  \kern-1pt\rddots & \kern4pt\rddots &  \\
        f_{n-1} & \kern1pt\rddots & 0 &  \\
        1& &  &
      \end{array}
    \right]\!\!
 \left[
      \begin{array}{ccc}
        g_{0} & \ldots & g_{n-1} \\
        &  \ddots &\vdots \\
        0 & &  g_0
      \end{array}
    \right]
- 
\left[
      \begin{array}{cccc}
        g_{1} & \kern-3pt\ldots & g_{n-1} & \kern2pt 0 \\
        \vdots &  \kern-1pt\rddots & \kern4pt\rddots &  \\
        g_{n-1} & \kern1pt\rddots & 0 &  \\
        0& &  &
      \end{array}
    \right]\!\!
 \left[
      \begin{array}{ccc}
        f_{0} & \ldots & f_{n-1} \\
        &  \ddots &\vdots \\
        0& &  f_0
      \end{array}
    \right].
\end{equation}
For deriving an analogous formula for $H_A ^{-1}$, using the notations
of Section~\ref{subsec:constantterm}, we first recall that 
 $H=\tilde{{\mathcal K}}_u {\mathcal K}_v$ and 
$H_A = \tilde{{\mathcal K}}_u A {\mathcal K}_v$. Multiplying~(\ref{eq:linsysf})
on the left by $\tilde{{\mathcal K}}_u A \tilde{{\mathcal K}}_u^{-1}$ gives 
\begin{equation} \label{eq:linsysHA}
H_A \,[
f_0, f_1, \ldots, f_{n-1}]^T
= - 
[h_{n+1}, h_{n+2}, \ldots, h_{2n}]^T.
\end{equation}
We also notice that
$$
H_A H ^{-1}= \left( {\mathcal K}_u ^{-1} A ^T  {\mathcal K}_u\right)^T,
$$
and, using the action of $A ^T$ on the vectors $u ^T, \ldots, (A
^T)^{n-2}u ^T$, we check that $H_A H ^{-1}$ is the companion matrix
$$
H_A H ^{-1} = 
\left[
 \begin{array}{cccc}
        0  & 1 &   & 0\\
         \vdots &   & \kern4pt\ddots  & \\
        0 & \ldots & 0 & \kern8pt 1  \\
        -f_0 & -f_1  & \ldots & -f_{n-1}  
      \end{array}
\right].
$$ 
Hence the last column $[g_0^*, g_1 ^*,\ldots, g_{n-1}^*]$ of $H_A ^{-1}$ is 
the first column of $H^{-1}$ divided by
$-f_0$. Using~(\ref{eq:invH}) for determining the first column of $H
^{-1}$, we get 
\begin{equation} \label{eq:lastcolHA}
[g^*_0, g^*_1, \ldots, g^*_{n-1}]^T = -\frac{g_0}{f_0}[f_1, \ldots,
f_{n-1},1]^T
+[g_1, \ldots,
g_{n-1},0]^T.
\end{equation}
Applying~\citep[Theorem\,3.1]{LaChCa90}, now  
with~(\ref{eq:linsysHA}) and~(\ref{eq:lastcolHA}),
we obtain
\begin{equation} \label{eq:invHA}
 H_A^{-1}= \left[
      \begin{array}{cccc}
        f_{1} & \kern-3pt\ldots & f_{n-1} & \kern2pt 1 \\
        \vdots &  \kern-1pt\rddots & \kern4pt\rddots &  \\
        f_{n-1} & \kern1pt\rddots & 0 &  \\
        1& &  &
      \end{array}
    \right]\!\!
 \left[
      \begin{array}{ccc}
        g^*_{0} & \kern-1pt\ldots & \kern-1pt g^*_{n-1} \\
        &  \kern-1pt\ddots &\kern-1pt\vdots \\
        0& &  \kern-1pt g^*_0
      \end{array}
    \right]
- 
\left[
      \begin{array}{cccc}
        g^*_{1} & \kern-3pt\ldots & g^*_{n-1} & \kern2pt 0 \\
        \vdots &  \kern-1pt\rddots & \kern4pt\rddots &  \\
        g^*_{n-1} & \kern1pt\rddots & 0 &  \\
        0& &  &
      \end{array}
    \right]\!\!
 \left[
      \begin{array}{ccc}
        f_{0} & \ldots & f_{n-1} \\
        &  \ddots &\vdots \\
        0 & &  f_0
      \end{array}
    \right].
\end{equation}

From~(\ref{eq:invH}) and~(\ref{eq:invHA})
we see that computing $\sigma _{k} (H ^{-1})$ and $\sigma _{k-1} (H_A ^{-1})$,
for $0\leq
k \leq 2n-1$,
reduces to computing the anti-diagonal sums for a product of 
triangular Hankel times triangular Toeplitz matrices. 
Let 
$$
M = LR =
\left[
\begin{array}{cccc}
        l_{0} & l_1 & \ldots & l_{n-1}  \\
        l_1  &  \kern11pt\rddots & \kern4pt\rddots &  \\
        \vdots & \kern1pt\rddots & 0 &  \\
        l_{n-1} & &  &
      \end{array}
\right]
\left[
\begin{array}{cccc}
        r_{0} & r_1 & \ldots & \kern8pt r_{n-1}  \\
          &  \ddots & \kern4pt\ddots &\kern8pt r_{n-2}  \\
         &0  &  \kern8pt\ddots & \vdots \\
         & &  & \kern4pt r_0
      \end{array}
\right]. 
$$
We have 
\begin{equation} \label{eq:tmpm1}
m_{i,j} = \sum _{s =i-1}^{i+j-2} l_{s} r_{i+j-s -2},
~1\leq i+j-1 \leq n,
\end{equation}
and 
\begin{equation} \label{eq:tmpm2}
m_{i,j} = \sum _{s =i-1}^{n-1} l_{s} r_{i+j-s -2},
~n \leq i+j-1 \leq 2n-1.
\end{equation}
For $0 \leq k \leq 2n-2$, $\sigma _k(M)$ is defined by summing the
$m_{i,j}$'s such that $i+j-2=k$. Using~(\ref{eq:tmpm1}) we obtain 
$$ 
\begin{array}{ll}
\sigma _k (M) & = \sum _{i=1}^{k+1} m_{i,k-i+2} = \sum _{i=1}^{k+1}
\sum _{s=i-1}^{k} l_s r_{k-s}, \\
& = \sum _{s=0}^{k} (s+1) l_s r_{k-s}, ~0 \leq k \leq n-1,
\end{array}
$$ 
hence 
\begin{equation} \label{eq:prodsigma1}
(\sum _{s=0}^{n-1} l_s x ^{s+1})' (\sum _{s=0}^{n-1} r_s x ^{s})
\, \bmod x^n =  \sum _{k=0}^{n-1} \sigma _k(M) \,x^k.
\end{equation}
In the same way, using~(\ref{eq:tmpm2}) with $\bar{k}=k-n+2$, we have
$$ 
\begin{array}{ll}
\sigma _k (M) & = \sum _{i=1}^{n-\bar{k}+1} m_{i+\bar{k}-1,n-i+1} = \sum _{i=1}^{n-\bar{k}+1}
\sum _{s=i}^{n-\bar{k}+1} l_{s+\bar{k}-2} r_{n-s}, \\
& = \sum _{s=\bar{k}-1}^{n-1} (s+n-k) \,l_{s} r_{k-s} ,~n-1 \leq k \leq 2n-2,
\end{array}
$$ 
and
\begin{equation} \label{eq:prodsigma2}
(\sum _{s=1}^{n} r_{n-s} x ^{s})' (\sum _{s=0}^{n-1} l_{n-s-1} x ^{s})
\, \bmod x^n =  \sum _{k=0}^{n-1} \sigma _{2n-k-2}(M) \,x^k.
\end{equation}

It remains to  apply~(\ref{eq:prodsigma1}) and~ (\ref{eq:prodsigma2})
to the structured matrix products in~(\ref{eq:invH}) and~(\ref{eq:invHA}),
for computing the~$\sigma _k(H ^{-1})$ and~$\sigma _k(H_A ^{-1})$'s.
Together with the minimum polynomial $f=f_0 + \ldots + f_{n-1} x^{n-1} + x^n$,
let $g=g_0 +  \ldots + g_{n-1} x^{n-1}$ (see~(\ref{eq:linsyslastcol})),
and  $g^*=g^*_0  \ldots + g^*_{n-1} x^{n-1}$ (see~(\ref{eq:lastcolHA})).
We may now combine, respectively (\ref{eq:invH}) and (\ref{eq:invHA}), with (\ref{eq:prodsigma1}),
for obtaining 
\begin{equation} \label{eq:sigmaklowH}
f'g-g'f \bmod x^n =  \sum _{k=0}^{n-1} \sigma _k(H ^{-1}) \,x^k,
\end{equation}
and 
\begin{equation} \label{eq:sigmaklowHA}
f'g^*-(g^*)'f \bmod x^n =  \sum _{k=0}^{n-1} \sigma _k(H_A ^{-1}) \,x^k.  
\end{equation}
Defining also $\mbox{rev}(f)=1 +  f_{n-1} x + \ldots + f_0 x^n$,
$\mbox{rev}(g)=  g_{n-1} x + \ldots + g_0 x^n$,
and $\mbox{rev}(g^*)=  g^*_{n-1} x + \ldots + g^*_0 x^n$,
the combination of, respectively, (\ref{eq:invH}) and (\ref{eq:invHA}), with (\ref{eq:prodsigma2}),
leads to 
\begin{equation} \label{eq:sigmakhighH}
\mbox{rev}(g)'\mbox{rev}(f) - \mbox{rev}(f)'\mbox{rev}(g)
\bmod x^n =  \sum _{k=0}^{n-1} \sigma _{2n-k-2}(H) \,x^k,
\end{equation}
and 
\begin{equation} \label{eq:sigmakhighHA}
\mbox{rev}(g^*)'\mbox{rev}(f) - \mbox{rev}(f)'\mbox{rev}(g^*)
\bmod x^n =  \sum _{k=0}^{n-1} \sigma _{2n-k-2}(H_A) \,x^k.
\end{equation}

\begin{prop} \label{prop:computsigma}
Assume that the minimum polynomial $f$ and the Hankel matrices $H$ and
$H_A$ are given. The anti-diagonal sums  $\sigma _{k} (H ^{-1})$ and
$\sigma _k (H_A ^{-1})$, for 
$~0 \leq
k \leq 2n-1$, can be computed in ${\sf G}(n)+O({\sf M}(n))$ operations in $\K$.
\end{prop}

Using the approach of~\citet{BGY80} we know that 
computing the last column of $H^{-1}$ 
reduces to an extended Euclidean problem of degree $2n$.
Hence the polynomial $g$ is computed in 
${\sf G}(n)+O(n)$ operations.  From there, $g^*$ is
computed using~(\ref{eq:lastcolHA}).
Then, applying~(\ref{eq:sigmaklowH})-(\ref{eq:sigmakhighHA}) leads to
the cost $O({\sf M}(n))$.


\bibliographystyle{elsart-harv}

\newcommand{\Gathen}{\relax}\newcommand{\Christian}{\relax}

\end{document}